\documentclass[twocolumn,showpacs,preprintnumbers,amsmath,amssymb]{revtex4}
\usepackage{graphicx}
\usepackage{dcolumn}
\usepackage{bm}
\usepackage{amsmath}
\begin{document}

\preprint{}
\title{Manipulation of Majorana fermion, Andreev reflection and
Josephson current on topological insulators}
\author{Yukio Tanaka$^{1}$, Takehito Yokoyama$^{2}$ 
and Naoto Nagaosa$^{2,3}$}
\affiliation{$^1$Department of Applied Physics, 
Nagoya University, Nagoya, 464-8603,
Japan \\ 
$^2$ Department of Applied Physics, University of Tokyo, Tokyo 113-8656, Japan \\
$^3$ Cross Correlated Materials Research Group (CMRG), ASI, RIKEN, WAKO 351-0198, Japan 
}
\date{\today}

\begin{abstract}
We study theoretically charge transport properties 
of normal metal (N) / ferromagnet insulator (FI) / superconductor (S)
junction and S/FI/S junction formed on the 
surface of three-dimensional topological insulator (TI), 
where chiral Majorana mode (CMM) exists at FI/S interface. 
We find that CMM generated in  N/FI/S and S/FI/S junctions are 
very sensitively controlled by the direction of the magnetization ${\bm m}$ 
in FI region. 
Especially, 
the current-phase relation of Josephson current in S/FI/S junctions 
has a phase shift neither 0 nor $\pi$, which 
can be tuned continuously by the component of ${\bm m}$ 
perpendicular to the interface. 
\end{abstract}

\maketitle



%

%



The class of time-reversal (TR) symmetric insulators 
with the nontrivial topological properties has been 
proposed theoretically and discovered experimentally 
\cite{Kane,Zhang1,Konig}.
The hallmark of this insulator, i.e.,
topological insulator (TI), is the 
edge or surface channels, while the bulk states
are gapped and inert for the low energy 
phenomena. In the two-dimensional (2D) case, there appears the
helical modes at the edge of the sample, i.e., 
the pair of one-dimensional modes
connected by the time-reversal symmetry and
propagating in the opposite directions for opposite 
pseudo-spins~\cite{Zhang1}.
This is analogous to the chiral edge modes in the 
quantum Hall systems~\cite{Wen}.
Therefore, the TI in 2D case can be regarded as the
two copies of the quantum Hall systems for up and down 
pseudo-spins~\cite{Onoda}, 
and is often called the quantum spin Hall system.
In three-dimensional (3D) case, there are two class of TI, i.e., weak TI (WTI) and strong TI (STI), corresponding to the even (WTI) and odd (STI) 
number of the chiral Dirac fermions on the surface
\cite{FuKane3D}.
Since the even number of Dirac fermions can be paired to open the gap, those in WTI is fragile against the disorder and/or the interactions 
while they are robust in STI. 
The 2D quantum spin Hall system is adiabatically 
continued to the WTI when the weak interlayer 
coupling is tuned, while the STI has no analogue to 
the quantum Hall system, and is a genuine new state of matter. \par
The 2D chiral Dirac fermion on the surface of STI is 
protected by the bulk gap and its topological property. 
Therefore, it offers an interesting system to look for 
the 2D superconductivity with the Cooper pairs 
mediated by the bosonic excitations, e.g., 
phonons and excitons, in the STI. 
Fu and Kane studied the superconductivity
induced by the proximity effect to the surface of STI~\cite{Fu1}.
Considering the interface between the ferromagnetic 
insulator (FI) and conventional superconductor (S),
they predicted the appearance of the chiral 
Majorana  state  as an Andreev bound state ~\cite{Fu1}. 
We call this chiral Majorana mode (CMM), 
which has a dispersion along the interface 
while it is confined along the direction perpendicular to the interface. 
Detecting the Majorana fermions in terms 
of the interferometory has been proposed also \cite{Fu2,Beenakker1}. \par
%
%
The presence of CMM is predicted in 
the $p_{x} + ip_{y}$   chiral superconductors \cite{Majoranap}, 
e.g., Sr$_2$RuO$_4$. 
However the  control of the chiral domains and 
manipulation of the edge channels 
are experimentally difficult\cite{Maeno}. 
In addition, the $p_{x} + ip_{y}$ superconductivity 
is very fragile against the disorder.
Also, the CMM in the 
$^{3}$He and cold atoms  have been studied theoretically \cite{Volovik}, 
but they 
are neutral systems and charge transport is missing there.
Therefore, the present system offers unique opportunity to study
the quantum charge transport specific to CMM and its control, which is more promising to be realized 
experimentally. However, its theoretical studies have been limited
focusing on the detection of the Majorana fermion itself 
\cite{Fu2,Beenakker1,Beenakker2}.   \par
In this paper, we study the manipulation of the 
quantum transport properties associated with Majorana 
fermions at the interface of the superconductor(S) and 
ferromagnetic insulator (FI)  
generated on the surface on TI. 
Hereafter, since we concentrate on the STI, we denote TI instead of STI for 
simplicity. 
%
We show that the direction of the magnetization ${\bm m}$ 
can be used to control the Andreev reflection and Josephson 
current via the CMM 
generated in  N/FI/S and S/FI/S junctions, 
offering a unique
method for superconducting spintronics. \par
%
\begin{figure}[htb]
\begin{center}
\scalebox{0.8}{
\includegraphics[width=11.0cm,clip]{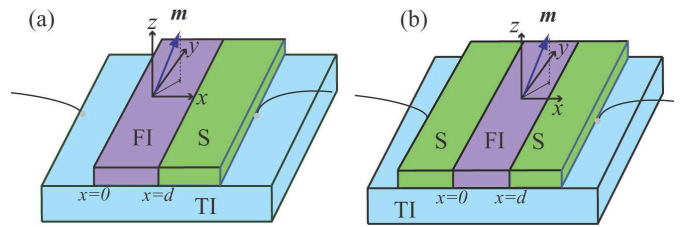}
}
\end{center}
\caption{
Schematic illustration of the junction. 
(a) Normal metal (N) / Ferromagnetic insulator (FI) / 
Superconductor (S) junction and 
(b) S/FI/S junction formed  on the surface of 
3D topological insulator (TI). The current is flowing on the surface of TI. 
}
\label{fig:1}
\end{figure}
We consider FI/S and S/FI/S structures formed on the surface of 
3D topological insulators as shown in Fig.1. 
We concentrate on the situation that TI below FI becomes ferromagnetic 
insulator due to the exchange coupling. Since surface state of TI is metallic, we can regard 
the configuration shown in Fig. 1(a) as  
normal metal (N)/ FI /S 
junction with N$(x<0)$, FI$(0<x<d)$, and S$(x>d)$. 
We also consider S/FI/S junction
as shown in Fig. 1(b) with S$(x<0)$, FI$(0<x<d)$, and S$(x>d)$. 
The interfaces N(S)/FI and FI/S locate at 
$x=0$ and $x=d$, respectively. 
We assume that the chemical potential in N and S are equal to each other. \par
The Hamiltonian of the surface state on TI 
is given by 
\begin{eqnarray}
\check{H}_{S} = \left( {\begin{array}{*{20}c}
{\hat H( {\bm k}) + \hat{M} } & {\hat \Delta } \\
{ - \hat \Delta ^*  } & { - \hat H^ * \left( 
{ - {\bm k}} \right) - \hat{M}^{*} } \\
\end{array}} \right)
\label{Hamiltonian}
\end{eqnarray}
with 
$\hat{H}(\bm{ k})=v_F(\hat{\sigma}_{x}k_{x}+\hat{\sigma}_{y}k_{y}) 
-\mu[\Theta(-x)+\Theta(x-d)]$,   
and 
$\hat{M}= {\bm m} \cdot \hat{\bm \sigma}\Theta(d-x) \Theta(x)$
with 
${\bm m} \cdot \hat{\bm \sigma}= m_{x}\hat{\sigma}_{x} +
m_{y}\hat{\sigma}_{y} + m_{z}\hat{\sigma}_{z}$. 
Here, $\mu$, 
$\hat{\bm{\sigma}}$, $v_F$, ${\bm m}$ denote
chemical potential, Pauli matrices, velocity, and 
magnetization (times the exchange coupling constant which we assume to be 1), 
respectively \cite{Fu1}. Note that ${\bm m}$ enters as an effective 
vector potential ${\bm A}$ of the electromagnetic field. 
The pairing symmetry of superconductor is assumed to be 
$s$-wave and $\hat{\Delta}$ is given as  
$\hat{\Delta} = i \hat{\sigma}_{y} 
\Delta \Theta(x-d)$ and  
$\hat{\Delta} = i \hat{\sigma}_{y} 
[\Delta \Theta(x-d) + \Delta \Theta(-x)\exp(i\varphi)]$  
for N/TI/S and S/TI/S junctions, respectively, 
where $\varphi$ denotes the macroscopic phase difference between 
left and right superconductor. 
In general, the magnitude of $\Delta$ is smaller than 
that of the bulk energy gap of superconductor deposited on TI 
due to the nonideal S/TI interface \cite{Fagas}. 
 \par
First, let us consider N/FI/S junction (Fig. 1(a)). 
A wave function of an electron injected from N with an injection angle $\theta$ is given as $\Psi_{T} =\exp(ik_{y}y)[\Psi_{N}(x)\Theta(-x) + 
\Psi_{FI}(x) \Theta(x)\Theta(d-x) + \Psi_{S}^{R}(x) \Theta(x-d)]$, 
where $k_{y}=k_{F}\sin\theta$ is the momentum parallel to the interface with  
$v_F k_{F}=\mu$. 
$\Psi_{N}(x)$, $\Psi_{FI}(x)$ and 
$\Psi_{S}^{R}(x)$ are given by 
\begin{eqnarray}
\!\!\Psi_{N}(x)\!\!\!&=& \!\!\!
[(\Psi_{in} \!+\! a \Psi_{hr})\exp(ik_{x}x) 
\!+\! b \Psi_{er} \exp(-ik_{x}x)] 
\label{wave1}
\\
\!\!\Psi_{FI}(x)\!\!\!\!&=&\!\!\!
[\Psi_{e1}\exp(-\tilde{\kappa}x) 
+ \Psi_{e2}\exp(\tilde{\kappa}^{*}x) 
\nonumber\\
&&+\Psi_{h1}\exp(\tilde{\kappa}x) 
+ \Psi_{h2}\exp(-\tilde{\kappa}^{*}x) ]
\label{wave2}
\\
\!\!\Psi_{S}^{R}(x) \!\!\!&=& \!\!\!
[ \Psi_{et} \exp(ik_{x}x) 
+ \Psi_{ht} \exp(-ik_{x}x) ] 
\label{wave3}
\end{eqnarray}
with 
$k_{x}=\sqrt{(\mu/v_F)^{2}-k_{y}^{2}}$, 
$\tilde{\kappa}_{x}=\kappa_{x} + im_{x}/v_F$, 
and 
$\kappa_{x}=\sqrt{m_{z}^{2} + (v_F k_{y} + m_{y})^{2}}/v_{F}$. 
The four component wave functions 
$\psi_{in}$, $\psi_{hr}$ and $\psi_{er}$
are given by $^T\psi_{in} =\left(
1,\exp(i\theta),0,0 \right)$, 
$^T\psi_{hr} =\left(0,0,1,-\exp(i\theta) \right)$, 
$^T\psi_{er} =\left(1,-\exp(-i\theta),0,0 \right)$, 
with $\exp(i\theta)=(k_{x}+ ik_{y})/k_{F}$. 
Other wave functions  
$\psi_{e1}$, 
$\psi_{e2}$, 
$\psi_{h1}$, 
$\psi_{h2}$, 
$\psi_{et}$, 
and 
$\psi_{ht}$ are obtained by solving eq. (\ref{Hamiltonian}) in a similar way. 
\par
%
%
The coefficients of Andreev reflection $a$ and normal reflection $b$ 
are obtained by imposing the boundary condition 
$\Psi_{N}(x=0)=\Psi_{FI}(x=0)$,  and 
$\Psi_{FI}(x=d)=\Psi_{S}^{R}(x=d)$. 
Then, the angle resolved 
tunneling conductance for injection angle $\theta$ 
is obtained by standard way  as  
$\sigma_{S}(\theta)=1 + \mid a \mid^{2} -\mid b\mid^{2}$. 
The normalized angle averaged 
tunneling conductance $\sigma$ by its value in the normal state 
is given by \cite{TK95}
\begin{eqnarray}
\sigma &=& \frac{\int^{\pi/2}_{-\pi/2} \sigma_{S}(\theta) \cos\theta d\theta }
{\int^{\pi/2}_{-\pi/2} \sigma_{N} \cos\theta d\theta}
\label{average}
\\
\sigma_{S}(\theta)&=&
\frac{\sigma_{N} [1 + \sigma_{N} \mid \Gamma \mid^{2}
-(1-\sigma_{N})\mid \Gamma \mid ^{4}] }
{ \mid 
1 + (1 - \sigma_{N})\exp(i\gamma) \Gamma^{2} \mid^{2} }
\label{angle}
\end{eqnarray}
with 
$\exp(i\gamma) = [m_{z}\cos\theta + i(\mu \sin\theta + m_{y})]/
[m_{z}\cos\theta - i(\mu \sin\theta + m_{y})]$ and 
$\Gamma=\Delta/(E + \sqrt{E^{2} -\Delta^{2}})$. 
$\sigma_{N}$ denotes the transparency of the junction in the normal state 
given by 
$\sigma_{N}=1/(\cosh^{2}(\kappa_{x} d) + \tan^{2}\theta 
\sinh^{2}(\kappa_{x}d)k_{y}^{2}/\kappa_{x}^{2})$. 
For $\sigma_{N} \rightarrow 0$, 
the denominator of $\sigma_{S}(\theta)$ becomes zero at $E=E_{b}$
\begin{eqnarray}
E_{b}=-\frac{ \Delta (\mu \sin\theta + m_{y}) {\rm sgn}(m_{z})}
{\sqrt{(\mu \sin\theta + m_{y})^{2} + \cos^{2}\theta m_{z}^{2}}}. 
\label{CMM}
\end{eqnarray}
This condition coincides with the 
formation of CMM   
at the FI/S interface with semi-infinite. 
CMM and Andreev reflection are strongly related to each other 
and $\mid a \mid$=1 is satisfied at $E=E_{b}$ independent of $\sigma_{N}$. 
It is noted that the sign of $E_{b}$ is changed by reversing the 
direction of magnetization $m_{z}$. 
$E_{b}$, $\sigma_{S}(\theta)$ and $\sigma$ are independent of  
$m_{x}$.
%
%
%
As a special limit of $m_{z}=\mu$ and $m_{y}=0$, 
this formula includes the 
case of chiral  $p_{x} + i {\rm sgn}(-m_{z}) p_{y}$-wave  
superconductor
where $E_{b}$ is reduced to be 
$E_{b}= - \Delta \sin\theta {\rm sgn}(m_{z})$ \cite{Yamashiro} 
although the given pair potential is full gap $s$-wave. 
It is remarkable that the sign of $m_{z}$ corresponds to the 
chirality of the CMM, which can be controlled by
the direction of magnetization of FI. 
%
Here, we focus on how the above CMM influences the 
bias voltage $V$ dependence of 
conductance in N/FI/S junctions with $E=eV$.  
As shown in Fig. 2, the resulting $\sigma$ has a 
zero bias conductance peak (ZBCP) 
originating from the peak of $\sigma_S(\theta)$ at $E=E_b$. 
%
As seen from the left panel of Fig. 2, the 
slope of the curve of $E_{b}$ 
around $E_{b}=0$ ($\theta=0$) becomes gradual with the decrease 
of $\mu/m_{z}$. Then, the contribution around $\theta=0$ 
becomes significant in the integral of numerator in 
eq. (\ref{average}) 
and the resulting height of the ZBCP is enhanced with the 
decrease of the magnitude of $\mu/m_{z}$ as shown in  
the right panel. \par
\begin{figure}[htb]
\begin{center}
\scalebox{0.8}{
\includegraphics[width=8cm,clip]{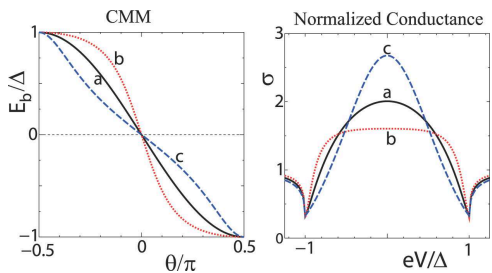}
}
\end{center}
\caption{(Color online) 
Left Panel: Chiral Majorana mode (CMM) energy level $E_{b}$ 
as a function of the incident angle $\theta$. 
Right Panel: 
Normalized tunneling conductance $\sigma$ in N/FI/S junctions. 
$m_{z}d/v_F=1$ and $m_{y}/m_{z}=0$. 
a: $\mu/m_{z}=1$, b: $\mu/m_{z}=2$ and c: $\mu/m_{z}=0.5$. 
}
\label{fig:2}
\end{figure}
The presence of $m_{y}$ changes $E_{b}$ drastically 
since $E_{b}(\theta)=-E_{b}(-\theta)$ does not hold any more.
As shown in the left panel of Fig.3, in 
the presence of $m_{y}$, $E_{b}$ does not become zero 
any more for $\theta=0$. For $m_{y}/m_{z}=0.4$, the domain 
of $\theta$ with negative $E_{b}$ is larger than that of 
positive one. 
The resulting $\sigma$ has a peak at negative bias 
voltage as seen from curve $b$ in the right panel of Fig. 3. 
On the other hand, for $m_{y}/m_{z}=-0.4$, the domain 
of $\theta$ with positive $E_{b}$ is larger than that 
of negative one as seen from curve $c$ in the left panel. 
Then the resulting $\sigma$ has a peak at positive bias voltage 
as seen from curve $c$ in the right panel  of Fig. 3.  
As shown above, 
CMM, Andreev reflection and resulting $\sigma_{S}(\theta)$ are 
critically controlled by ${\bm m}$. 
The present CMM is significantly different from that in the 
non-centrosymmetric superconductor, where CMM appears 
as helical edge modes \cite{NCS}. 
In the present case, one of the spin-split bands 
is missing compared with the non-centrosymmetric 
superconductors. 
\par
\begin{figure}[htb]
\begin{center}
\scalebox{0.8}{
\includegraphics[width=8cm,clip]{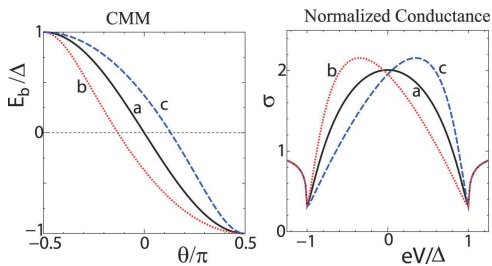}
}
\end{center}
\caption{(Color online) 
Left Panel: CMM energy level $E_{b}$.
Right Panel: 
$\sigma$ in 
N/FI/S junctions. $m_{z}d/v_F=1$ and $\mu/m_{z}=1$. 
a: $m_{y}/m_{z}=0$, b: $m_{y}/m_{z}=0.4$ 
and c: $m_{y}/m_{z}=-0.4$. 
}
\label{fig:3}
\end{figure}
Next, we focus on the Josephson 
current in S/FI/S junction (Fig. 1(b)). 
Since the magnitude of the pair potential of the 
left and right superconductors are equal to each other, it is possible to 
evaluate the Josephson current by using the CMMs   
formed in S/FI/S junctions \cite{Tanaka97}. 
The CMMs can be obtained 
as in the case of N/FI/S junction with wave functions of Eq. (\ref{Hamiltonian}) under a proper boundary condition. 
The resulting Josephson current $I$ can be obtained as
\begin{eqnarray}
I &=& \frac{2e}{\hbar}\frac{\partial F}{\partial \varphi},
\ \ F=-k_{B}T \sum_{\bm k} {\rm log} 
[2\cosh(\frac{E_{J}}{2k_{B}T})],
\label{FreeJ}
\\
E_{J}&=&
\sqrt{\sigma_{N} \cos^{2}(\varphi/2 -\delta) 
+ (1 - \sigma_{N})(E_{b}/\Delta)^{2}}\Delta
\label{CMMJ}
\end{eqnarray}
with $\delta=m_{x}d/v_F$ and temperature $T$. 
The appearance of $\delta$ in the current-phase relation 
stems from the fact that 
$\tilde{\kappa}_{x}$ in 
eq. (\ref{wave2}) 
becomes complex number in the presence of $m_{x}$.
The position of the 
present CMMs in S/FI/S junctions is 
given by $\bar{E}_{b}=\pm E_{J}$. 
The expression of $\bar{E}_{b}$ can be explained by the 
hybridization of two CMM formed at the left S/FI interface 
and right FI/S interface. 
The formula of $\bar{E}_{b}$ [eq. (\ref{CMMJ})] is 
very general including several preexisting cases.
As a special limit with $E_{b}=\Delta$, and $m_{x}=0$, 
$\bar{E}_{b}$ is reduced to 
$\bar{E}_{b}=\pm \sqrt{\cos^{2}(\varphi/2)+(1-\sigma_{N})
\sin^{2}(\varphi/2)}\Delta$, 
which corresponds to the Andreev bound state in 
conventional  S/non-magnetic insulator (NI)/S junction \cite{Golubov}. 
If we choose $E_{b}=0$, and $m_{x}=0$, the bound state formula in 
$d_{xy}$($p_{x}$)-wave /NI/$d_{xy}$($p_{x}$)-wave  junction 
is reproduced \cite{Golubov}. 
To understand the $\theta$ and $\varphi$ dependence of $\bar{E}_{b}$ ($E_{J}$) 
more intuitively, we plot $E_{J}$(absolute value of $\bar{E}_{b}$) in Fig. 4. 
$ E_{J} $ is an even function of $\theta$ for any 
$m_{x}$. For $m_{x}=0$, $E_{J}$ is symmetric with respect to $\varphi=0$ (Fig. 4(a)). 
However, for $m_{x} \neq 0$, the 
resulting $E_{J} $ is no more symmetric around $\varphi=0$ 
as shown in  Figs. 4(b) and 4(c). 
This fact seriously influences the Josephson current $I$  
given by 
\begin{equation}
eIR_{N}=\frac{\sin(\varphi-2\delta)\int^{\pi/2}_{-\pi/2} 
d\theta \frac{\pi \Delta^{2}\tanh(E_{J}/2k_{B}T)\sigma_{N}\cos\theta}
{2E_{J}}}{\int^{\pi/2}_{-\pi/2} 
d\theta \sigma_{N} \cos\theta}
\label{Josephson}
\end{equation}
with resistance in the normal state $R_{N}$. 
\begin{figure}[htb]
\begin{center}
\scalebox{0.8}{
\includegraphics[width=8cm,clip]{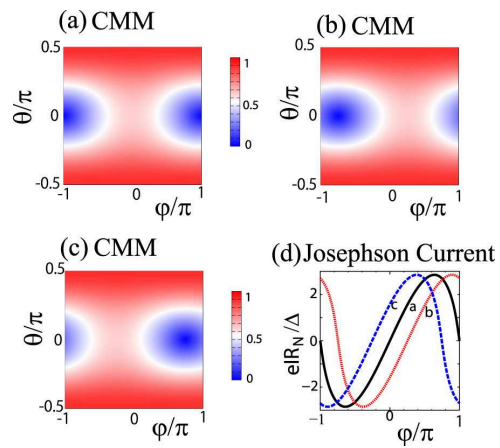}
}
\end{center}
\caption{(Color online) 
Contour plot of CMM energy level $E_{J}$ 
as a function of $\theta$ and $\varphi$ for (a)$m_{x}=0$, 
(b)$m_{x}=0.4m_{z}$, (c)$m_{x}=-0.4m_{z}$. 
The resulting 
Josephson current in S/FI/S junctions 
is plotted  in (d) with a: $m_{x}/m_{z}=1$, b: $m_{x}/m_{z}=0.4$ 
and c: $m_{x}/m_{z}=-0.4$. 
In all panels, we choose 
$m_{z}d/v_F=1$, $\mu/m_{z}=1$ and $m_{y}/m_{z}=0$. $T=0.05T_{C}$ with 
transition temperature $T_{C}$.}
\label{fig:4}
\end{figure}

Differently from $\sigma$ for the Andreev reflection, $I$ is almost independent  of $m_{y}$ because the contributions to the change in $I$ almost cancel 
between the two CMMs with opposite chiralities. 
On the other hand, $m_{x}$ seriously influences $I$
as the effective vector potential which directly enters
into the phase of the Josephson coupling $\varphi$.
Here, the absence of the small factor of $e/c$,
which reduces the coupling to the magnetic field, 
makes the tuning of the current-phase relation much easier. 
Reflecting the $\varphi$ dependence of 
$E_{J}$, 
the resulting $I$ 
becomes zero at neither $\varphi=0$ nor $\varphi=\pm \pi$
(curves $b$ and $c$ in Fig. 4(d)) due to the presence of the 
phase shift $2\delta$.  
Up to now, there have been many studies about Josephson current in 
ferromagnet junctions \cite{Fogel}, where
the value of phase shift becomes neither 0 nor $\pi$ in usual cases. 
The intermediate phase shift has been predicted 
in a few cases even in the absence of Majorana Fermion, 
i)spin-singlet $s$-wave / spin-triplet superconductor 
junction \cite{Asano}, 
ii)even-frequency / odd-frequency superconductor 
junction \cite{Tanaka-Ueda}, and 
iii)ferromagnet junction with spin-singlet  
$s$-wave superconductor in the presence of the 
spin-flip scattering or spin-orbit coupling \cite{Eschrig,Buzdin}. 
It is remarkable that this  anomalous current-phase relation 
appears simply controlling one magnetization vector 
without using unconventional pairing in the present model.  
It is also noted that the resulting 
current-phase relation can be continuously tuned by the change of $m_{x}$ 
similar to the control of the magnetization vectors 
at the interfaces of ferromagnet junction \cite{Eschrig}.
We hope this anomalous current-phase relation will be 
detected experimentally in SQUID. 
\par
In conclusion, we have studied the
charge transport properties 
of normal metal (N) / ferromagnet insulator (FI) / superconductor (S) junction and S/FI/S junction formed on the 
surface of three-dimensional (3D) topological insulator, where 
chiral Majorana mode (CMM) exists at FI/S interface. 
We have found that CMM generated in  N/FI/S and S/FI/S junctions are 
controlled by the direction of the magnetization ${\bm m}$ 
in FI region very sensitively. 
Since metallic surface state of 3D topological insulator has been observed experimentally, our proposed structure will be accessible in the 
near future \cite{BiSb}. 
Our results give a guide to innovate novel direction of  
superconducting spintronics.  \par
Note added: After a submission of thie paper, 
a realization of the 
Majorana fermion in superconductor / semiconductor heterostructure has been predicted in the presence of interplay of spin-orbit coupling and exchange field 
\cite{Sau}. 

This work is partly supported by the
Grant-in-Aids from under the Grant No.\ 20654030, 
No. 19048015 and No. 19048008 from 
MEXT, Japan, and NTT basic research laboratories.
T.Y. acknowledges support by JSPS.


\end{document}